# On the physics and technology of gaseous particle detectors

P Fonte[1,2], V Peskov[3]

[1] Laboratório de Instrumentação e Física Experimental de Partículas, Coimbra, Portugal
[2] Instituto Superior de Engenharia de Coimbra, Coimbra, Portugal
[3] CERN, Geneva, Switzerland.

E-mail: fonte@coimbra.lip.pt

**Abstract.** Despite an already long and fruitful history, gaseous elementary-particle detectors remain today an important mainstay of high-energy and nuclear physics experiments and of radiation detection in general. In here we briefly describe some of the gaseous detector's main technologies and applications, along with some unsolved gas-discharge physics aspects of practical relevance.

## 1. Introduction

Introduced by H.Geiger in 1908 [1] under the form of cylindrical single-wire counters working in a saturated discharge mode (Geiger mode), gaseous detectors evolved to a large number of specialized configurations, adapted to detect all forms of radiation with energy above 4eV.

Highlights of this technology, when compared with other types of radiation detectors, include the economic coverage of large areas and volumes, low specific mass and very good time and position resolution.

The operation is based on the Townsend avalanche, triggered by the primary ionization created by the impinging radiation and developing in specialised "amplification regions", where a large electric field is applied. In some cases the avalanche is followed by subsequent discharge regimes, which provide further charge amplification. The electric impulse thus developed is collected on patterned electrodes that provide position information and recorded by electronic means. In some applications it is advantageous to detect also the gaseous scintillation light emitted by the avalanches.

The first modern electronically-readout gaseous detector was the Multiwire Proportional Chamber (MWPC) introduced by G.Charpak et al. in 1968 [2]. Charpak was awarded the 1992 Nobel Prize for this one and other developments in gaseous particle detectors.

Usually MWPC is constituted by three parallel planes of wires: an anode plane and two cathode planes. Typically the distance between the planes is a few mm and the wire's pitch is about 1 to 3 mm. Depending on the electronic readout scheme, aimed at locating the avalanches, the wires on both cathode planes can be perpendicular to the anode wires or wires of one of the cathode plane could be parallel to the anode wires and the other one perpendicular. An important feature of the MWPC is that an avalanche created near a particular anode wire will induce in this wire a negative signal whereas on the surrounding anode and cathode wires the signal polarity will be positive and their amplitude will rapidly drop with the distance from the avalanche. This allows to determine the bidimensional position of the avalanches: in the direction perpendicular to the anode wires with an accuracy often better than the wire's pitch and in the direction along the wire with accuracy better than 100 µm.

Another type of classic detectors that find today important developments and applications include the remarkable "Pestov counter" [3] and its more practical successor, the Resistive Plate Chamber (RPC) [4].

## 2. Modern detector technologies

For the last four decades gaseous detectors have enjoyed a continuous flow of new inventions and innovations that have continually extended their range of applicability and there is a strong drive to further pursue this route [5]. Recent fundamental innovations include micropattern detectors [6], timing Resistive Plate Chambers (RPCs) [7] and the negative-ion drift chamber [8].

Micropattern detectors, of which there are today many variants [9], are characterized by very small amplification regions (sometimes below 50 µm) finely distributed on a surface and manufactured by modern microelectronics technology. This allows a very good, isotropic, 2-D position resolution in digital readout mode[1] over a finely segmented readout electrode. Due to the small distance between the electrodes they have good time resolution and counting rate capability. Moreover, microelectronic technology allows direct integration of the gas amplification structures with the signal microelectronic readout.

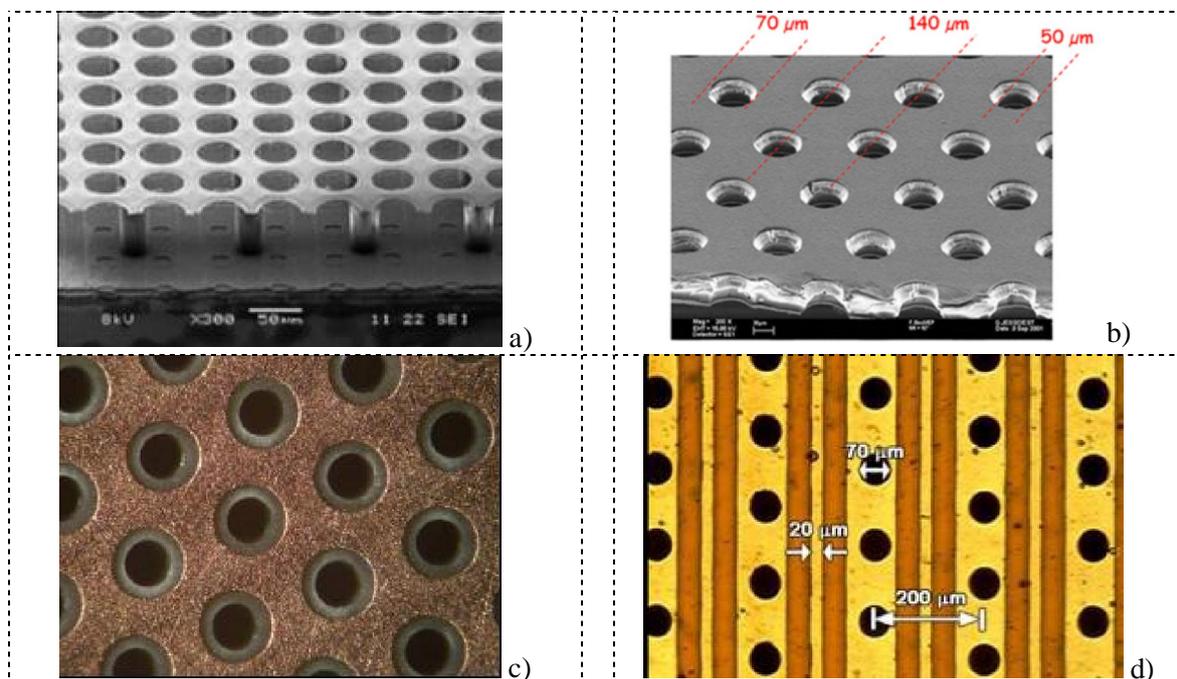

**Figure 1.** Depiction of some modern micropattern gaseous detectors: a) InGRID MICRO MEsh GAS chamber (MICROMEGAS); b) Gaseous Electron Multiplier (GEM); c) Thick GEM; d) Micro Hole Strip Plate (MHSP).

Some of the most prominent designs of micropattern detectors are shown in figure 1. In figure 1a it is depicted a photograph of a so called InGrid MICROMEGAS: a metallic micromesh placed 50 to

---

[1] Recording only the location of the readout electrode segments with signal above a certain threshold, as opposed to performing a charge measurement on each segment and deducing the avalanche position from the charge distribution.

100 μm above a pixelized (pixel size 50×50 μm) anode board in which each pixel is connected to a microamplifier and the signal from these amplifiers are readout by a quite sophisticated electronics allowing not only to determine the position of the avalanches with an accuracy better than 50 μm , but also to measure the avalanches timing and perform various other measurements. The so-called Gas Electron Multiplier (GEM) is shown in figure 1b. As it can be seen from the photographs, there is some similarity between InGrid MICROMEGAS and GEM; however avalanches in GEM are strictly confined inside the holes (see [3] for more details). Finally, there are some practical modifications of the GEM principle. One is the thick GEM (figure 1c) and another the microhole strip plate (figure 1d) [10]. Thick GEM is a robust GEM-like detector which operates at 10 times higher gain than GEM and can withstand even powerful spark discharges. The microhole strip chamber allows to combine pre-amplification in holes with the subsequent multiplication around fine strips.

It is evident that the fine structure of the micropattern detectors electrodes renders them fragile and they (except for the thick GEM) can be easily destroyed by occasional sparks, almost unavoidable during high-gain or long-term operation. For this reason, the latest tendency in micropattern detector development is the implementation in their design of resistive electrodes (e.g. [11]). This does not increase the maximum achievable gain of the detectors but makes them spark-protected.

Timing RPCs are made from flat glass plates that define thin (200 to 300 μm), but mechanically accurate, gas gaps. These detectors, offering an outstanding time resolution of 50 ps [12], opened the way for very large area (e.g. [13]) time-of-flight detectors for relativistic particles. Additionally, timing RPCs equipped with fine-pitch readout strips offer high position resolution (~30μm in digital mode) and were already successfully used for radiography (see section 3.4).

The negative-ion drift chamber - a drift chamber operated in slightly electronegative gas mixtures - apart from being a very original idea, opens the way for very precise, albeit slow, Time Projection Chambers [8].

## 3. Main current applications

In here we will point out a representative list of present-day applications and refer the reader to the relevant documents for further information on the underlying detector technologies. It is clear that a comprehensive review of all past or even present applications of gaseous detectors is impossible in the form of a short article. Our choice is certainly influenced by our professional background; many other equally relevant examples were certainly left unmentioned.

### 3.1. High Energy Physics

Gaseous detectors in one form or another are used in the vast majority of High Energy Physics experiments. An exhaustive list would include essentially all such experiments.

Important examples include the larger experiments at CERN's [14] Large Hadron Collider (LHC) accelerator, ATLAS [15] and CMS [16], where these are extensively used on the huge outer detector layers, covering thousands of square meters. A remarkable exception is ATLAS' Transition Radiation Tracker [17], which is the central large-volume tracker of the experiment, doubling up as an X-ray detector for electron identification at high energies by transition radiation[2]. The gaseous detector structures employed include single-wire counters (ATLAS' "straw tubes" [17], CMS' "monitored drift tubes" [18]), MWPCs (ATLAS' "cathode strip chambers" [19]) and RPCs on the muon trigger systems of both experiments [18]-[19].

It should be also mentioned the central high-rate tracker of the COMPASS experiment at CERN [20], based on GEMs [6].

---

[2] The phenomenon by which a relativistic particle crossing a sharp boundary between media with different permittivity may emit a soft X-ray. Electrons are much more likely to do so than other, heavier, particles.

## 3.2. Nuclear Physics

Also at CERN LHC, the ALICE [21] high-energy heavy-ion experiment is strongly based on gaseous detectors (figure 2). Technologies include: MWPCs at the "cathode pad chambers" of the muon arm [22], as a sophisticated large area photodetector for RICH[3] at the high momentum particle identification system [23] and as a soft X-ray detector for the transition radiation detector [24]; a large time projection chamber [25]; RPCs in the muon arm trigger [22] and in the time of flight system [13].

At GSI [26] the FOPI [27] and HADES [28] experiments are also essentially based on gaseous detectors, including most of the techniques mentioned above, plus a "jet chamber", "multidrift chambers" and a "pre-shower" detector using an MWPC working on the "self-quenched streamer" mode.

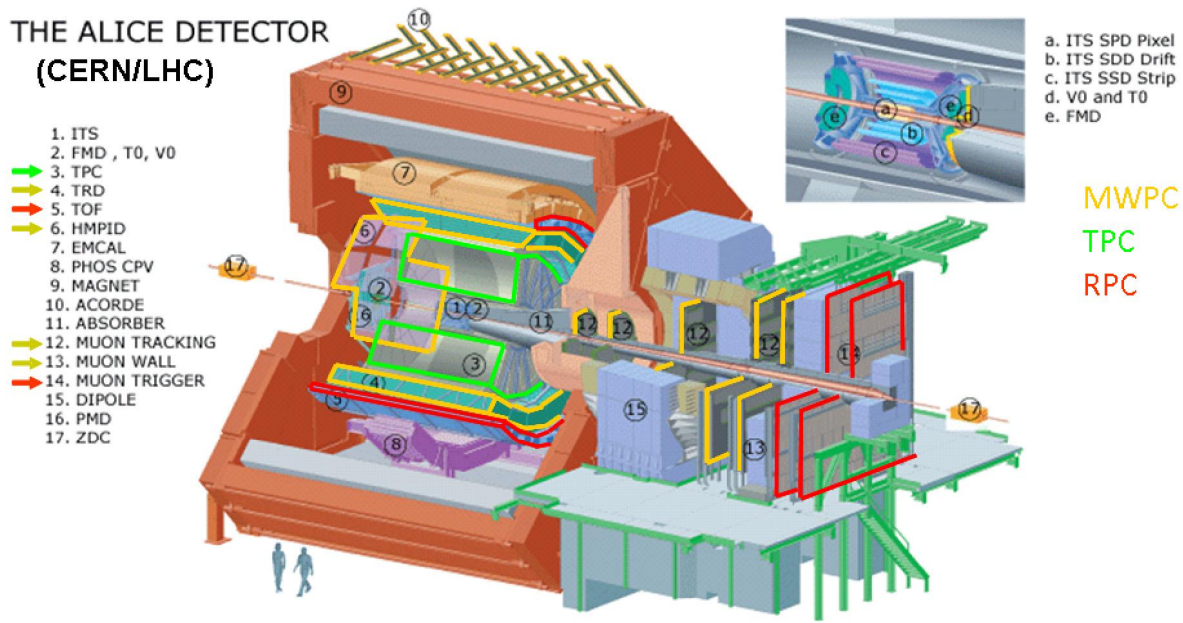

**Figure 2.** Representation of the ALICE detector at CERN [21]. The gas-based sub-detectors are outlined and their respective technologies indicated. It can be appreciated that this modern experiment strongly relies on gaseous detectors.

## 3.3. Astronomy

In Astronomy, gaseous detectors find application in two specialized niches: X-ray spectroscopy and X-ray polarimetry.

The first application is based on the Gas Proportional Scintillation Counter [29] and has been used in observational satellites [30]-[31].

The second application, the Gas Pixel Polarimeter [32], is a recent development based on a GEM directly coupled to a very finely pixelized readout chip. It relies on the fact that the direction of the photoelectron emerging from an X-ray photoelectric interaction is related to the photon's polarization. The concept has been selected for the ESA X-ray observational satellite XEUS [33].

Furthermore, a large-area charged particle detector, based on RPCs, is now deployed in a cosmic-ray observatory [34].

---
[3] Ring Imaging CHerenkov: the detection of the photons, forming a ring, emitted by fast particles via the Cerenkov effect, yields information about the velocity of a relativistic particle.

## 3.4. Medicine and Biology

Gaseous detectors have been proposed and prototypes developed for digital radiography [35]-[38] and Positron Emission Tomography [39]-[41]. Examples can be seen in figure 3.

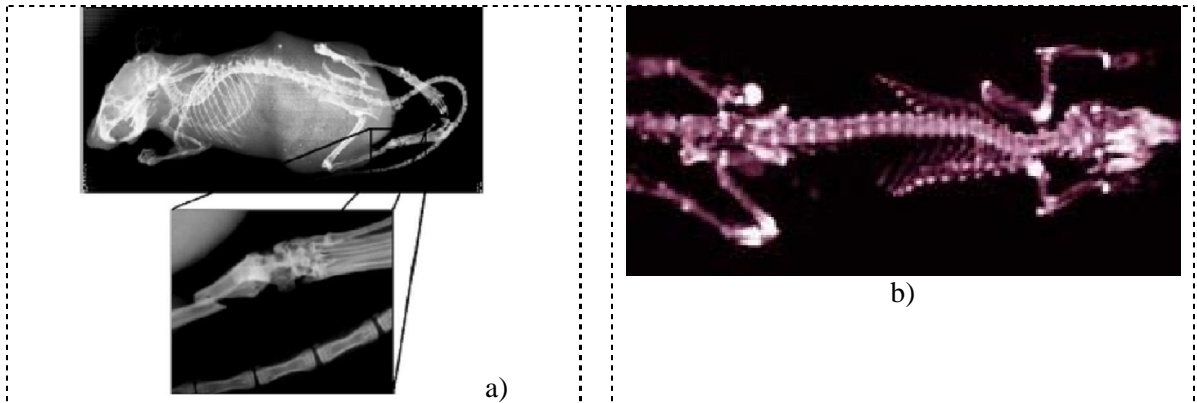

**Figure 3.** Examples of images of mice taken with gaseous detectors: a) digital radiography [38]; b) Positron Emission Tomography [42].

## 3.5 Plasma diagnostics

Gaseous detectors, especially MWPC, were extensively used in plasma diagnostics: for visualization of its UV and X-ray radiation (see, for example, [43]-[45]); measurements of the X-ray spectrum (for example [46]-[49]); measurements of arc atomic temperature [53]; studies of near-electrode phenomena (for example [50]). Recently, reports appeared on the application of micropattern gaseous detectors for plasma diagnostics [51]-[52]. This list is certainly incomplete, but it gives a flavour on the ongoing very fruitful applications of gaseous detectors for various plasma studies

## 4. Breakdown characteristics

One important practical limitation of gaseous detectors arises from the phenomenon of breakdown, leading to sparks, which limits the maximum charge attainable in a single avalanche and the maximum counting rate at a certain gas gain. Both aspects can be appreciated in figure 4, which shows the limiting curves for many detector types in the space of gain vs. counting rate density.

### 4.1. Gain-induced breakdown

The maximum gain limit at low counting rate density is relatively well understood. It was recognized to correspond to the avalanche charge that causes the avalanche's own space-charge generated electric field to become comparable to the applied field close to the anode, allowing the formation of a positive "streamer" [58]. This limit is historically known as the "Raether limit" [59], amounting roughly to $10^8$ electrons in the avalanche. However, in the context of gaseous detectors, particularly in micropattern detectors, this figure is frequently smaller by several orders of magnitude as the avalanches may be of very small spatial extent.

However, as the theory of streamer discharges is today not yet fully developed [60] and the knowledge of crucial aspects, such as non-local photoionization characteristics, is deficient even for the most studied gas mixture (air), it cannot be said that this very important aspect of detector operation is completely understood.

Detectors may contribute to the development of the theory of streamers by providing some specific insights originated from a rich body of empirical experimentation:

- essentially all non-toxic gas mixtures were tried at one or another of the many detector geometries, including strongly electronegative mixtures containing sulphur hexafluoride;
- some detectors (RPCs) work in a regime where the avalanches are deeply affected by space-charge, which is only possible if the development of the positive streamer is somehow suppressed, albeit only partially as small streamers are still observable;
- in RPCs there is experience as to which additives enhance or suppress the streamer development, Ar being an important example of a reinforcing additive;
- pre-ionization mechanisms [61] can be ruled out, as these would be readily apparent in the observed signals.

For the large majority of these very diverse experimental conditions the development of positive streamers is always observed (an important exception is the "Geiger mode" that occurs around thin anode wires). This universality seems to be at odds with the complex interplay of the details of photoemission, photoabsorption and photoionization spectra and yields that must be considered for every component of a gas mixture and for their avalanche-generated reaction products in any theory of non-local photoionization. It may be that a more general mechanism is at work.

For instance, considering pure simple hydrocarbons at atmospheric pressure, conditions on which most, if not all, detectors work well, one readily concludes (e.g. [62]) that close to their ionization thresholds the photoabsorption cross-sections are close to $5 \times 10^{-17}$ $cm^{-2}$, corresponding to an absorption length of the order of 10 μm. It is hard to see how such drastic photoabsorption may allow a viable feedback mechanism for positive streamer growth, even without making any considerations about emission characteristics. Of course, one may consider also the ionization of reaction products, but in here the discussion will become much more complicated and likely it will be difficult to find the universal conditions that seem to be experimentally suggested.

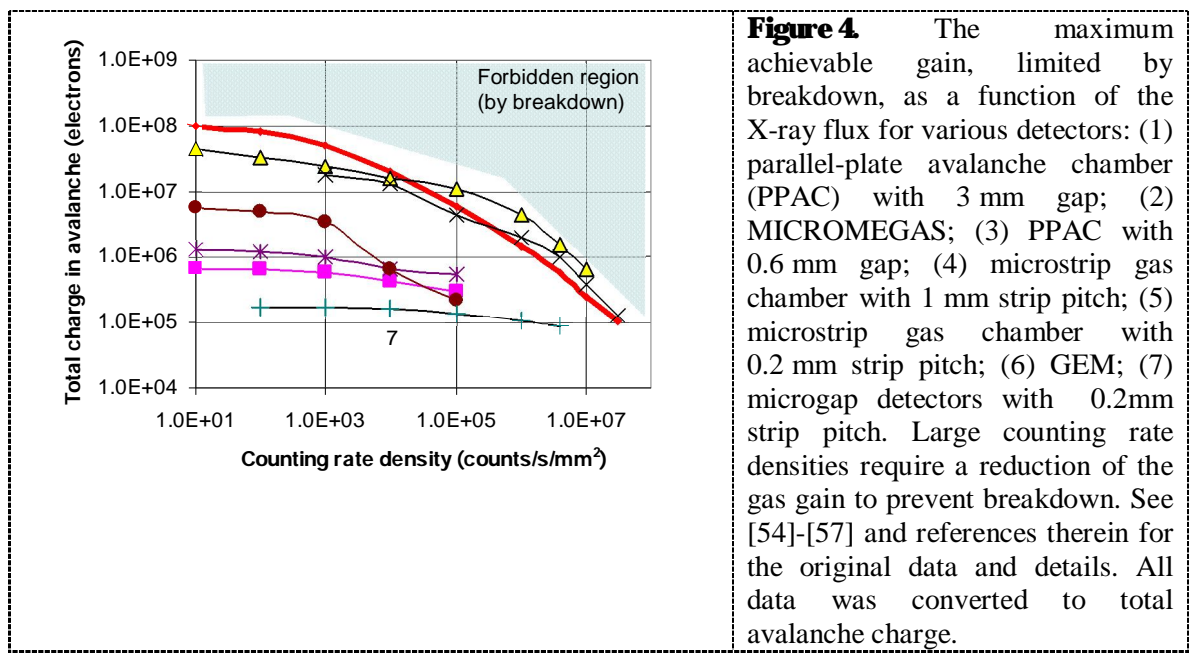

**Figure 4.** The maximum achievable gain, limited by breakdown, as a function of the X-ray flux for various detectors: (1) parallel-plate avalanche chamber (PPAC) with 3 mm gap; (2) MICROMEGAS; (3) PPAC with 0.6 mm gap; (4) microstrip gas chamber with 1 mm strip pitch; (5) microstrip gas chamber with 0.2 mm strip pitch; (6) GEM; (7) microgap detectors with 0.2mm strip pitch. Large counting rate densities require a reduction of the gas gain to prevent breakdown. See [54]-[57] and references therein for the original data and details. All data was converted to total avalanche charge.

*4.2. Rate-induced breakdown*

The second aspect – rate-induced breakdown – is even less known and its physical origin remains today unclear. For example, as follow from figure 4, the maximum achievable gain of most gaseous detectors (except wire-based) drops with the counting rate. A possible explanation of this phenomenon based on avalanche overlapping in time and space is presented in [63]. However, there are other phenomena simultaneously observed besides simple avalanche overlapping. In figure 5 we present some observations related to these phenomena.

In Parallel Plate Avalanche Chamber (PPAC), at low gas gain and under strong irradiation it is observed that there is a "preparation period", up to a few seconds, over which the detector average current increases spontaneously, eventually leading to breakdown (figure 5a). A time-resolved observation of the period just before breakdown suggests that the current increases either via an increase of the rate of individual avalanches (avalanche "jets"- figure 5b) or by a steady current (figure 5d).

At somewhat larger gains the preparation period is much shorter and the detector may not spontaneously recover from the breakdown event, sparking repeatedly until the applied voltage is reduced (figure 5c). Again, there may be avalanche "jets" just before breakdown (not shown).

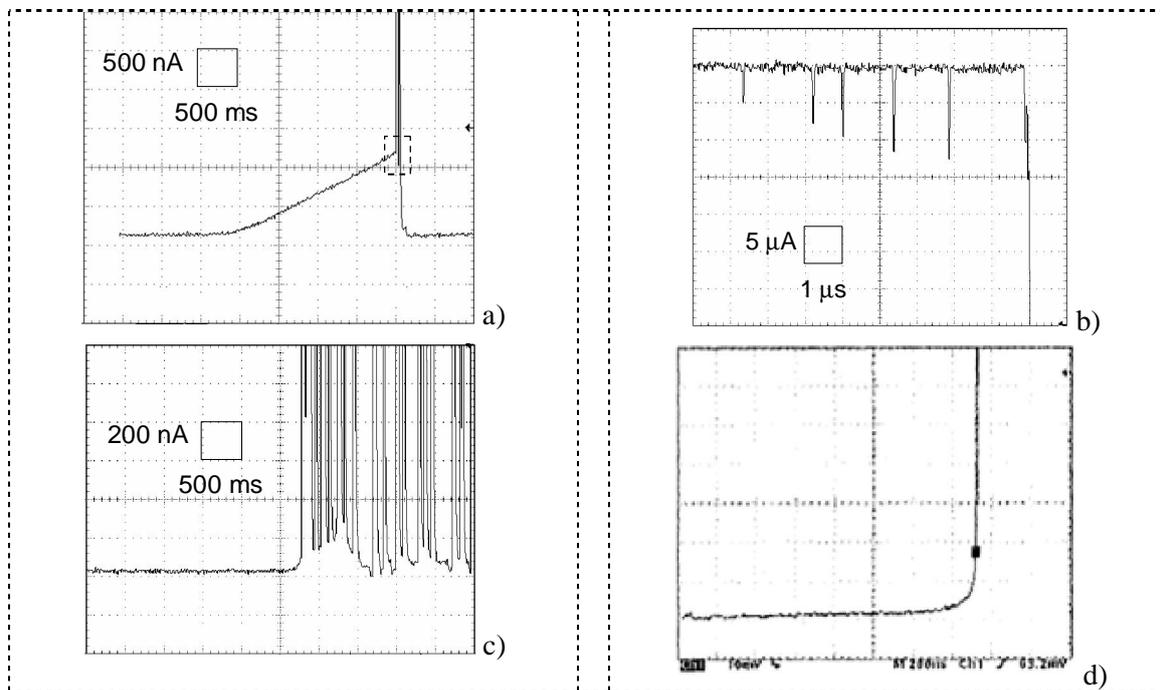

**Figure 5.** Average current (left) and time-resolved current (right) in a 3 mm gap PPAC under strong irradiation at a low gas gain (top) and at a higher gas gain (bottom) [64] [65]. a) the detector average current increases spontaneously ("preparation" period), eventually leading to breakdown. b) A time-resolved observation of the period just before breakdown suggests a strong increase of the rate of individual avalanches (avalanche "jets"). c) At higher gas gains the transient period is shorter. The detector never recovers from the breakdown episode and sparks repeatedly ("memory" effect). d) high-rate breakdown may also be triggered by a short-term, but steady, current growth without visible "jets".

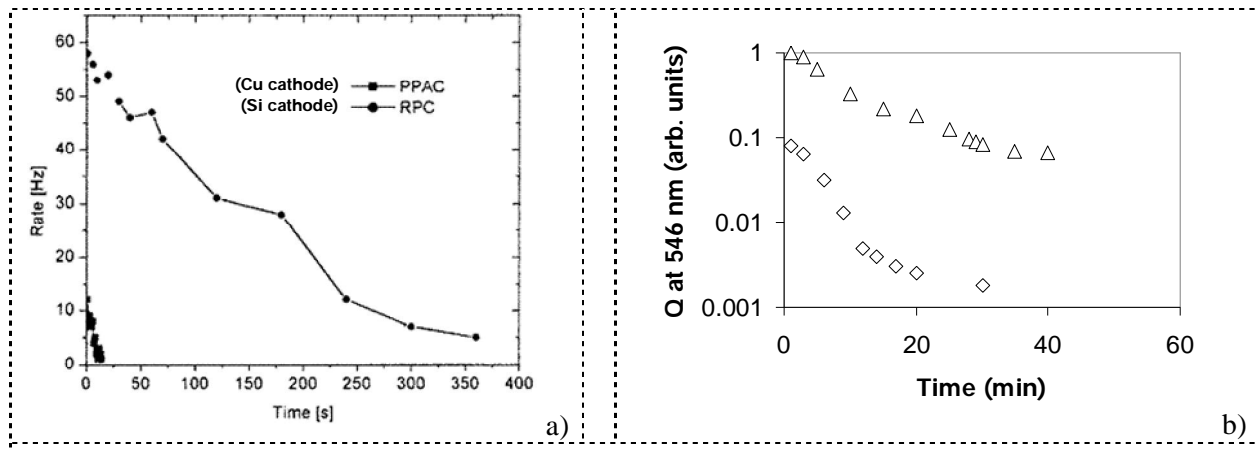

**Figure 6.** Some experimental data on "cathode excitation" effects that may be on the origin of the post-breakdown "memory" effect. a) After strong irradiation there is an excess of dark count rate, dependent upon the cathode material [65]. b) For many minutes after a breakdown episode there are strong changes of the cathode's quantum efficiency (Q) for visible light, both for Cu (rhombus) and CsI (triangles) photocathodes [63].

Typically the detector keeps a "memory" of the breakdown episode and the applied voltage cannot be taken back to the original value before many hours have passed. In figure 6 we show some experimental data on "cathode excitation" effects that may be on the origin of the "memory" effect. After strong irradiation there is an excess of dark count rate[4] and after a breakdown episode the cathode's photosensitivity is much enhanced. These effects decay in time on the scale of many minutes and seem to depend on the cathode material.

## 5. Acknowledgements

This work was performed in the framework of the RD51 collaboration and financed by the EU and FEDER via the Fundação para a Ciência e Tecnologia (FCT) project CERN/FP/83524/2008.

---

[4] Spontaneously generated avalanches, most likely from electrons emitted by the cathodes.